**Visual anemometry: physics-informed inference of wind for renewable energy, urban sustainability, and environmental science**

John O. Dabiri[1,2]*, Michael F. Howland[3], Matthew K. Fu[1], Roni H. Goldshmid[1]*

1. Graduate Aerospace Laboratories, California Institute of Technology, Pasadena, CA 91125, USA

2. Mechanical and Civil Engineering, California Institute of Technology, Pasadena, CA 91125, USA

3. Civil and Environmental Engineering, Massachusetts Institute of Technology, Cambridge, MA 02139 USA

*Authors for correspondence: jodabiri@caltech.edu, ronigoldshmid@gmail.com

**Abstract**

Accurate measurements of atmospheric flows at meter-scale resolution are essential for a broad range of sustainability applications, including optimal design of wind and solar farms, safe and efficient urban air mobility, monitoring of environmental phenomena such as wildfires and air pollution dispersal, and data assimilation into weather and climate models. Measurement of the relevant multiscale wind flows is inherently challenged by the optical transparency of the wind. This review explores new ways in which physics can be leveraged to "see" environmental flows non-intrusively, that is, without the need to place measurement instruments directly in the flows of interest. Specifically, while the wind itself is transparent, its effect can be visually observed in the motion of objects embedded in the environment and subjected to wind—swaying trees and flapping flags are commonly encountered examples. We describe emerging efforts to accomplish visual anemometry, the task of quantitatively inferring local wind conditions based on the physics of observed flow-structure interactions. Approaches based on first-principles physics as well as data-driven, machine learning methods will be described, and remaining obstacles to fully generalizable visual anemometry will be discussed.

**Key Points**

- Atmospheric winds near the Earth's surface mediate a variety of essential physical, chemical, and biological processes at length-scales ranging from millimeters to kilometers across the globe.

- A better understanding of wind dynamics can enable more efficient renewable energy technologies, more accurate monitoring and modeling of weather and climate, and more rapid adoption of new technologies such as urban air mobility.

- Visual anemometry is an emerging technique that aims to infer quantitative estimates of wind speed and direction based on visual observations of associated flow-structure interactions, such as swaying trees and flapping flags.

- Physics-based models of flow-structure interactions and data-driven models that leverage machine learning and artificial intelligence have both demonstrated initial proofs-of-concept that site-specific visual anemometry is feasible.

- Generalized visual anemometry—a technique that does not require calibration measurements or a priori collection of training data—will likely depend on the discovery of new physical principles that govern the diversity of environmental structures exposed to wind.

I. Introduction

The fate of life on Earth depends on macroscopic physical processes that are nonetheless imperceptible to the naked eye. Specifically, the movement of air masses at local scales mediates essential gas exchanges between the atmosphere and the terrestrial and aquatic ecosystems that lie underneath[1–8]. This flow of air is also a principal means of transportation for life ranging from bacteria[9] and seed spores[10–14] to animals that migrate seasonally across the globe[15–17]. Engineering technologies with the promise to protect those same ecosystems are also dependent on the wind. The functional reliance of technologies like wind turbines is a straightforward example[18–20]; however, it may be less appreciated that the performance of solar energy farms is also determined by local wind conditions[21,22]. Text Box 1 provides further discussion of the diverse roles of wind flows in environmental sustainability applications.

Given this broad and important role of the wind for current and future environmental sustainability, it is remarkable that we have relatively few tools available to quantify the wind at the length and time scales relevant to many of the applications identified above. Such measurements are inherently limited by the optical transparency of the air. To date, the most common solutions to this limitation require introducing an engineered, physical object into the flow whose interaction with the wind can be detected visually as a qualitative indicator (e.g., a windsock or wind vane[23]) or alternatively, by converting the physical interaction of the object and the wind into a calibrated, quantitative signal (e.g., a cup anemometer or light detection and ranging (LiDAR) system[24–26]). Measurements using these approaches are all fundamentally constrained by the requirement that the measurement device must be located in close proximity to the measurement domain of interest.

This review explores an emerging alternative with the potential to enable multiscale, spatiotemporally resolved measurements of the wind by leveraging trillions of wind indicators already covering most of the land on Earth. These indicators include naturally occurring structures, such as the estimated three trillion trees on land[27], as well as engineered structures, such as the millions of kilometers of electrical power lines[28]. Because none of these objects is perfectly rigid, they move in response to local wind conditions in ways that could potentially be used to infer incident wind speed and direction. We call this technique *visual anemometry*, reflecting the opportunity to quantify local winds based solely on visual measurements at arbitrarily far, line-of-sight distances away from the region of interest.

The concept of visual anemometry as a qualitative tool was first popularized in 1805 with the eponymous scale introduced by Francis Beaufort, a British naval officer, to standardize assessments of the effect wind loading on ship sails. Subsequently adapted to wind over land, the Beaufort scale categorizes wind speed according to its qualitative effect on objects in the environment, from the gentle fluttering of leaves in low winds to the swaying motion of entire trees in high winds[29,30]. Companion scales have subsequently been developed to categorize higher wind speeds, such as the Fujita scale for tornadoes[31]; time-averaged wind speeds over longer periods are categorized by the Putnam-Griggs index[32]. The distinct objective of this review is to explore the convergence of physics and data science to achieve visual anemometry that is quantitative in its assessment of the observed flow-structure interactions and generalizable to any structures that exhibit a visible response in the presence of wind. The content of this review is complemented by prior surveys of flow anemometry more generally[33], as well as reviews of the fluid mechanics of urban canopies[34], plant canopies[13,35], forest canopies[36], and aquatic canopies[37].

Section II begins with an introduction to the relevant physics governing flow-structure interactions of the type expected to occur in wind flows. The remainder of the section reviews current approaches toward visual anemometry, while section III highlights remaining showstoppers to successful realization of this method. We conclude the review by identifying diverse ways in which the physics community can contribute their disciplinary expertise to the development of this emerging field.

II. Visual Anemometry

II. A. Principles of flow-structure interactions

II.A.1. Physics of vortex-induced vibration

Relative motion between a solid body and a surrounding fluid (such as that illustrated in Figure 1a) will create lift and drag forces (henceforth $F_L$ and $F_D$, respectively) that act on the body. These two forces act in the transverse and streamwise directions of the fluid, respectively, with magnitudes given by

$$F_L = \frac{C_L}{2} \rho U^2 A, \tag{1}$$

and

$$F_D = \frac{C_D}{2} \rho U^2 A, \tag{2}$$

where $\rho$ is the fluid density, $U$ is the incoming fluid speed, $A$ is the relevant body area, and $C_L$ and $C_D$ are the dimensionless lift and drag coefficients whose magnitude in an incompressible flow depends on the body's relevant Reynolds number ($Re$), shape, and orientation in the flow. Per convention, the Reynolds number is defined as $Re = \rho U L / \mu$, where $L$ is the relevant body length scale and $\mu$ is the dynamic viscosity of the fluid. While $C_L$ and $C_D$ will be nearly constant and of order 0.1 to 1 for rigid, bluff bodies at $Re \gg 1$, these coefficients can change and often decrease considerably with increasing flow speed, as will be discussed later. Lift and drag forces are responsible for momentum exchange between the fluid (e.g., surrounding air) and the body. When the body motion is coincident with one or both forces, it will extract kinetic energy from the fluid.

A simple example of this energy transference (or harvesting) occurs when a bluff body is placed in a steady (i.e., time-independent) flow. Across a wide range of flow conditions, these bodies generate an unsteady (i.e., time-varying) wake characterized by periodic vortex shedding—the formation of spatially compact regions of rotating fluid downstream of the body—at formation frequency $f_o$. This vortex formation results in a spatially uneven pressure distribution on the body. For a body with a single degree of freedom, such as an elastic cantilever that is allowed to move in the transverse direction (see Figure 1b), this unsteady forcing will cause the structure to respond by oscillating with so-called vortex-induced vibrations (VIV). When the forcing frequency, $f_o$, from the vortex shedding approaches the natural frequency, $f_N$, of the structure, the dynamics of the two systems can become coupled in a state of synchronization or "lock-in." This resonant state is characterized by a significant transfer of kinetic energy from the fluid to the structure resulting in large amplitude oscillations, i.e., comparable to the characteristic length scale of the body cross-section. This mechanism is responsible for phenomena such as "singing" wires and for the notable failure of the Ferrybridge cooling tower failures[38] in England. VIVs are one of many flavors of flow-induced vibrations, along with aeroelastic flutter instabilities[39] and galloping[40,41], which result from forcing due to unsteady pitching. Each of these response modes is a visually perceptible indication of the local wind conditions.

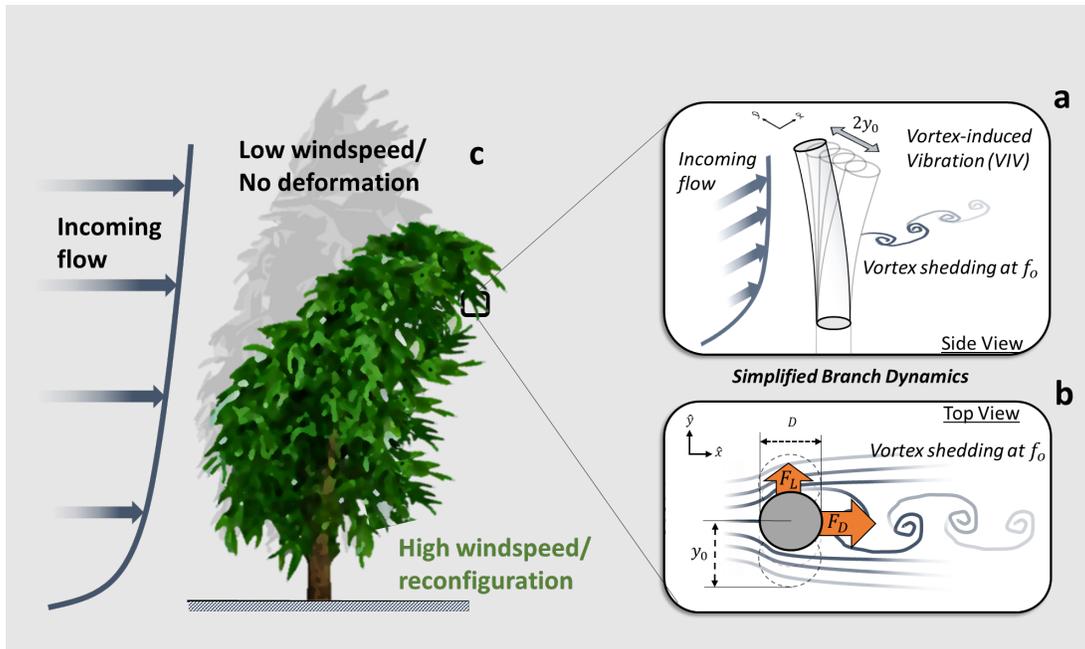

**Figure 1.** Physics of flow-structure interactions. Slender objects, such as the branches of a plant, can undergo vortex-induced vibrations in a direction transverse to the incoming flow (a), due to vortex shedding and associated lift and drag forces (b). In high-speed flows, flexible structures such as vegetation can undergo reconfiguration (c), which reduces proportional increases in aerodynamic forces from the wind.

II.A.2. Effects of flexibility and reconfiguration

For relatively rigid structures, the scaling in equations (1) and (2), especially the quadratic dependence on wind speed, adequately describes the behavior of aerodynamic forces exerted by the wind. However, many naturally occurring structures, especially plants, are highly flexible and thus can deform considerably under forcing from external fluid flow (Figure 1). Consequently, the flexural rigidity of the body plays an important role in determining the fluid forcing on many bodies[42]. Because of this flexibility, many plants will reconfigure their cross-section area and become more streamlined in higher speed flows, allowing them to experience a drag force with sub-quadratic dependency on flow speed[43]. This dependency can be expressed as $F_D \propto U^{2+V}$, where $V$ is the Vogel exponent[42,44–49]. Values of the Vogel exponent $V < 0$ capture the deviation from the canonical, inertial scaling relation due to reconfiguration. For example, $V = -1$ indicates a regime in which drag scales near linearly with velocity. Depending on the mechanics of the reconfigurability, system-specific parameterizations have been used to quantify reconfiguration in previous studies (e.g., see a recent review[42]). As discussed in the following sections, a goal of visual anemometry is to infer the relationship between structural response and incident wind without an *a priori* model of reconfiguration dynamics.

II.A.3. Effects of biological adaptation

The preceding discussion of reconfiguration implicitly assumes that the reconfiguration of the structure exposed to wind is reversible, i.e., that the original configuration of the structure is recovered when wind loading is removed. While this is typically true for engineering structures, provided that they are not stressed beyond the limits of elastic deformation, vegetation can exhibit a more complex response to wind loading. Specifically, many plants exhibit structural remodeling—at a cellular level—in response to wind loading that leads to different equilibrium geometries of the vegetation over time[50,51]. Hence, the structure of vegetation can encode the time history of previous exposure to wind loading. If not properly accounted for, this adaptation can potentially confound efforts toward visual anemometry, as two plants of the same species but with different wind loading histories could respond differently to incident wind, e.g., if one has grown stiffer due to previous exposure to the wind.

Conversely, by recognizing the occurrence of this phenomenon, a type of adaptation known as thigmomorphogenesis[52], additional information regarding the wind conditions can be beneficially leveraged in the task of visual anemometry. For example, the long-term deformation of trees has previously been exploited for the estimation of average annual wind speeds[32] and even for siting of wind turbines[53]. Hence, knowledge of the phenotypic plasticity of a given vegetation species (e.g., its capacity for thigmomorphogenesis) along with measurement of its equilibrium structure can complement visual observations of the instantaneous flow-structure interactions. The following section focuses on the physics governing the latter dynamic processes; data-driven strategies to incorporate information regarding tree species and wind history (e.g., equilibrium structure) are discussed in Sections II.B.2 and III.D.

II. B. Current approaches for visual anemometry

II.B.1. Physics-based methods

II.B.1.i Dynamics-based methods

Physical objects that are both geometrically slender and mechanically stiff have proven most amenable to direct, first-principles application of flow physics to deduce a quantitative relationship between object motion and incident wind speed. In these cases, the aerodynamic force of the wind on the structure can be estimated as $F_W \approx \bar{p}A$, where $p$ is the dynamic pressure exerted by the wind on the windward face of the structure, the overbar indicates a spatiotemporal average, and $A$ is the projected area of the corresponding surface. As described in Section II.A.1, the dynamic pressure is linearly proportional to the air density and quadratically proportional to the incident wind speed; hence,

$$F_W \propto \rho \bar{U}^2 A. \qquad (3)$$

The structural response to small deformations can be estimated by assuming that the elastic restoring force, $F_E$, is linearly proportional to the structure deflection:

$$F_E \approx \kappa \delta, \qquad (4)$$

where $\delta$ is the structural deflection and $\kappa$ is the elastic constant, which depends on the structure geometry and material properties. For cantilevered, slender objects such as tree branches, plant stalks, or blades of grass, the tip deflection due to spatially uniform wind loading can be modeled using linear Euler-Bernoulli beam theory, i.e.,

$$\delta \approx \frac{fL^4}{8EI}, \qquad (5)$$

where $f$ is the applied force per unit length $L$, $E$ is the elastic Young's modulus of the material comprising the structure, and $I$ is the area moment of inertia. Comparing equations (4) and (5) above, the corresponding elastic constant is

$$\kappa \approx \frac{8EI}{L^3}. \qquad (6)$$

The balance of aerodynamic and elastic forces, $F_W$ and $F_E$, respectively, provides a relationship between observed structural deflection and incident wind speed:

$$\bar{U} \approx \sqrt{\frac{8EI\delta}{\rho AL^3}} . \qquad (7)$$

Each of the parameters on the right-hand side of equation (7) can be estimated from visual observation of the structure, with the exception of the Young's modulus of the material. Ref. 54 demonstrated the use of single-point calibration to determine the unknown material property. Alternatively, computer vision techniques can potentially be used to deduce the likely material properties based on libraries of environmentally observed structures and their known material properties[55,56].

An important limitation of methods based on the preceding analysis is the necessary occurrence of a non-zero mean (i.e., time-averaged) structural deflection due to the incident wind. As described in Section II.A.1 above, the vortex-induced vibrations experienced by many environmental structures, such as plants[57], can exhibit a mean deflection that is close to zero despite significant instantaneous deflections. The oscillatory motion of electrical power lines under wind loading is another common example; other engineered structures, such as telephone poles and radio antennae, can also exhibit nearly symmetric structural oscillations in a direction perpendicular to the incident wind[28].

Ref. 58 showed that the dynamic motions associated with transverse or streamwise structural oscillations can also be used to estimate wind speeds, albeit using a conceptual framework different from the quasi-steady force balance that leads to equation (7). In this case, the dynamics of the flow-structure interaction are modeled as a damped harmonic oscillator:

$$F_W(t) \approx m\frac{d^2\delta}{dt^2} + \lambda\frac{d\delta}{dt} + \kappa\delta, \tag{8}$$

where $m$ and $\lambda$ are the structure mass and damping coefficients, respectively, and the last term on the right-hand side of the equation is the elastic response analyzed previously. In principle, any time-dependent wind forcing can be represented as a superposition of harmonic forcings[59] at a spectrum of frequencies:

$$F_W(t) = \frac{a_0}{2} + \sum_{n=1}^{N}(a_n \cos nt + b_n \sin nt) \tag{9}$$

where $a_n$ and $b_n$ are constants, and the summation includes $N$ modes sufficient to approximate the time-dependence of the incident wind. For harmonic forcing at a single frequency $\omega$, i.e., $F_W(t) = F_0 \sin \omega t$, equation (8) has the steady-state solution[58]

$$\delta(t) = \frac{F_0}{\kappa}\left[\frac{1}{(1-\beta^2)+(2\zeta\beta)^2}\right][(1-\beta^2)\sin(\omega t) - 2\zeta\beta \cos(\omega t)], \tag{10}$$

where $\beta$ is the ratio of the forcing frequency to the natural frequency of the structure, and

$$\zeta = \frac{\lambda}{\sqrt{4\kappa m}}. \tag{11}$$

Inspection of this steady-state solution indicates that the amplitude of the structural oscillations is directly proportional to the amplitude of the wind forcing. Appealing to the relationship between wind speed and forcing in equation (3) above, and with additional algebraic manipulation, Ref. 58 shows that, for typical, low levels of atmospheric turbulence, i.e., if

$$I_u \equiv \frac{\sigma(U)}{\bar{U}} \ll 1, \tag{12}$$

then

$$\bar{U} \propto \sqrt{\frac{\sigma(\delta)}{I_u}}, \tag{13}$$

where $\sigma$ denotes the standard deviation. Field measurements demonstrated that this relationship captures the flow-structure interactions of five tree species representing a diversity of morphologies[58,60].

As with the preceding mean deflection model, visual anemometry based solely on the physics of the time-dependent structural response still requires a calibration measurement to determine the constant of proportionality in the relationship expressed by equation (13). This potentially limits visual anemometry to contexts in which one has *a priori* wind measurements using conventional anemometry techniques. To unlock the potential of visual anemometry for global coverage, especially in regions where conventional anemometry is inaccessible, dynamical models like those above may require augmentation with other approaches.

II.B.1.ii Energy-based methods

While the mechanical properties of a structure exposed to wind can be difficult to infer based on visual observation of its isolated flow-structure interactions, the presence of multiple identical structures could be leveraged to infer their common properties. Consider, for example, the wind incident on two trees aligned in the streamwise direction. The discussion in the preceding sections indicates that the kinetic energy of each tree is derived from the kinetic energy of the incident wind, e.g., for the upstream tree:

$$KE_{T_1} \approx \eta KE_{WIND}, \qquad (14)$$

where $KE_{T_1}$ is the kinetic energy of the upstream tree, $KE_{WIND}$ is the kinetic energy of the wind incident on the front of the canopy, and $\eta$ is a constant factor that quantifies the energy transfer from the wind to the trees. This factor captures the mechanical properties of the tree, e.g., its inertia, elasticity, and damping. A value $\eta = 0$ would indicate no energy transfer from the incident wind to the tree, whereas a value $\eta = 1$ would indicate the (unphysical) upper bound of perfect energy transfer from the wind to the tree. In practice, the maximum value of the energy transfer coefficient $\eta$ is likely much less than 1. For example, in steady incompressible flow, the maximum theoretical value is given by the Betz limit of 59.3%[61]. Unsteady flows can exhibit higher energy transfer efficiencies in theory, though empirical observations suggest values less than the Betz limit are typical[62].

If the second, downwind tree is set into motion solely by remaining kinetic energy in the wake of the first tree, i.e., $KE_{WAKE_1} \equiv KE_{WIND} - KE_{T_1}$, then we can estimate the kinetic energy of the second tree as

$$KE_{T_2} \approx \eta KE_{WAKE_1} \approx \eta(1-\eta)KE_{WIND} \quad (15)$$

To be sure, this approximation assumes that no energy was dissipated in the interaction with the upstream tree (i.e., negligible damping on the timescale of wind advection), and it assumes that the upstream wind does not also directly affect the dynamics of the second, downwind tree, e.g., via turbulent sweeps into the top or sides of the canopy[24,35,63,64] or a redistribution of the kinetic energy within its frequency spectrum[65]. This approximation depends inherently on the level of turbulence in the incident wind and on the surrounding topography.

We postulate that two trees (or other objects exposed to the wind) with identical structural properties are characterized by the same value of energy transfer efficiency $\eta$. With this ansatz, we can eliminate the unknown structural properties $\eta$ by comparing the relative magnitude of the motion of the two trees:

$$\frac{KE_{T_1}}{KE_{T_2}} \approx \frac{1}{1-\eta}, \quad (16)$$

or,

$$\eta \approx 1 - \frac{KE_{T_2}}{KE_{T_1}}. \quad (17)$$

The kinetic energy of each structure can be estimated as proportional to the square of its average component speeds. Hence,

$$\eta \approx 1 - \frac{\widehat{U_{T_2}^2}}{\widehat{U_{T_1}^2}}, \tag{18}$$

where the carat denotes a spatial average of the structure motion. The model in equation (18) above could be enhanced by incorporating more realistic functional dependencies of the parameter $\eta$, for example, to reflect possible sensitivity of the kinetic energy transfer efficiency to the wind speed (e.g., via structure reconfiguration as discussed in section II.A.2. above) and background turbulence levels. However, these additions would potentially re-introduce the need for local calibration measurements. Even in its current form, equation (18) illustrates the potential for canopies comprising an array of similar structures to be especially useful for visual anemometry.

An additional physical phenomenon that can influence the accuracy of visual anemometry using the preceding energy-based methods is the presence of "honami," i.e., waves of wind-induced structural displacement that can propagate through a canopy[66,67]. For canopy elements in close proximity, mechanical contact between upstream and downstream elements during wave propagation could lead to additional kinetic energy transfer between canopy elements. Because this transfer of kinetic energy can be two-way—with energy transferred from upstream canopy elements to downstream, or vice versa—the upstream energy transfer could lead to underestimation of the wind kinetic energy transferred to downstream canopy elements and thereby underestimate $\eta$. For a uniform canopy, this should manifest as a systematic bias in the measurements. If so, then this artifact should be straightforward for a data-driven method to compensate based on training data (see sections II.B.2. and III.D).

Each of the aforementioned physics-based methods is fundamentally limited by the fidelity with which two-dimensional visual observations of the vegetation motion can accurately quantify the actual three-dimensional canopy kinematics[47,68–70]. A top view of the canopy provides a projection of the two dominant wind directions (i.e., streamwise and cross-wind), assuming that vertical wind flows are negligible as in canonical horizontally homogeneous atmospheric surface layer flows[64]. Hence, visual anemometry from this perspective is less sensitive to out-of-plane canopy motion that would lead to underestimates of the canopy kinetic energy. Moreover, because the vegetation is cantilevered at the ground, the portion of the canopy that is visible from overhead will typically exhibit the most significant displacements. This feature of overhead measurements becomes especially important for visual anemometry conducted from distant vantage points such as aircraft or satellites (see section III.B. for further discussion). The overhead perspective is also especially useful for inference of wind direction, which can be a desired output of visual anemometry irrespective of quantitative measurements of wind speeds.

Despite these advantages of overhead views of flow-structure interactions, many data sets of interest will necessarily be collected from ground-level perspectives, where the canopy is viewed from the side. In these cases, visual anemometry can only capture the projection of the wind in the plane perpendicular to the optical axis. Moreover, the wind associated with the observed canopy motions will be primarily the wind at the lateral sides of the canopy, as that is the primary visible interface between the canopy and the surrounding wind from a side view. If the motion of the top of the canopy is visible from the side, then it may also be possible to estimate wind at the upper interface of the canopy and the wind. In that case, one can anticipate a vertical gradient with wind speeds increasing from the ground to the top of the canopy.

Ultimately, three-dimensional canopy tracking, e.g., via light detection and ranging (LiDAR) could obviate the need for these considerations, as 3D reconstruction of the canopy motion would eliminate the aforementioned projection errors.

II.B.2. Data-driven methods

The constants of proportionality needed to complete the physical relationships expressed in equations (7), (13), and (14) above depend on factors specific to the objects being visually observed, e.g., their inertia, stiffness, and damping. Non-intrusive measurement of these properties at the scale of individual environmental structures is challenging, if not impossible, particularly when those objects comprise a heterogeneous composite of multiple materials. A potential way forward is to leverage the fact that the trillions of environmental objects of interest globally can be classified into a set of material categories that is several orders-of-magnitude smaller in number. For example, building codes limit the set of allowable compositions of artificial structures to a relatively small number of engineered materials[71,72]. These materials could therefore be deduced in many cases from the external appearance of the structures. As another example, high-voltage power lines are typically composed of an aluminum core and polyethylene insulation, both of standard physical dimensions[73–76]. Hence, the material properties of such an environmental structure can be deduced as soon as it is categorized.

Naturally occurring structures such as vegetation present a greater challenge, given both the diversity of their physical makeup and the fact that structure inertia, stiffness, and damping depend non-trivially on factors such as age, health, moisture content, and the presence or absence of leaves, seeds, and symbiotic organisms. Notwithstanding this myriad of challenges, initial efforts toward data-driven visual anemometry have produced encouraging results[77–79]. For example, Ref. 77 trained a combined convolutional neural network (CNN) and long short-term memory (LSTM) network based on field observations of a magnolia tree and a cloth flag exposed to naturally occurring wind conditions over several weeks. It was postulated that the CNN learns to recognize key features of the objects exposed to the wind, e.g., tree branches and leaves, or geometric patterns on the flag. Concurrently, the LSTM was hypothesized to learn key temporal features of the object motion, e.g., recurring waving or flapping motions of the geometric patterns.

The trained neural network was subsequently tested on video clips of the same tree and flag that were not included in the training data set. This purely data-driven visual anemometry achieved measurements of the mean wind speed with errors comparable with the background turbulence fluctuations at the field site of approximately 1-2 m/s (Figure 2).

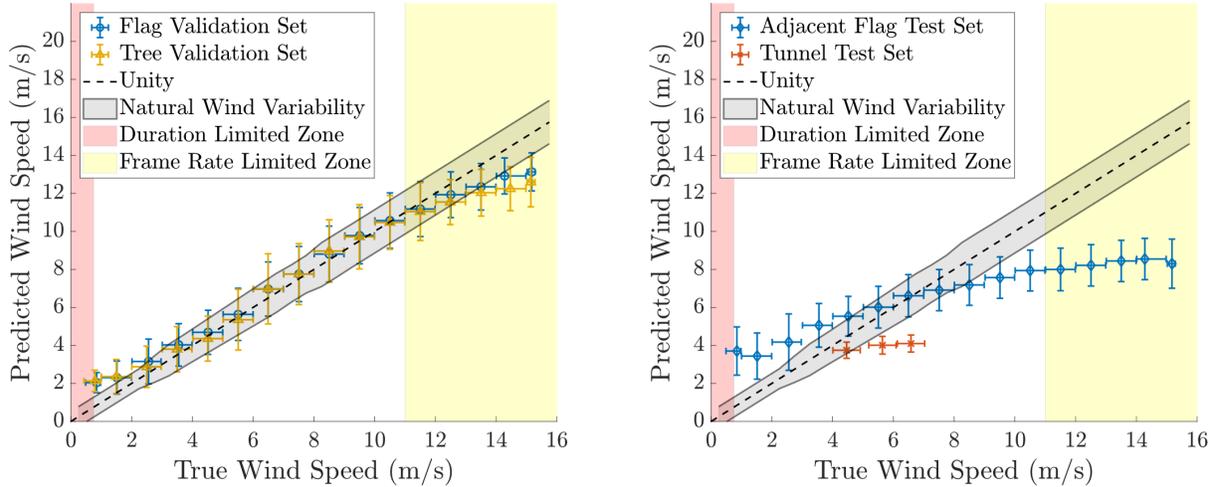

**Figure 2.** Data-driven implementation of visual anemometry based on measurements collected at a research field site and in a laboratory wind tunnel. A combined CNN-LSTM network successfully predicted the wind speed corresponding to new videos of the same structures included in the training data set (left panel). However, significantly lower sensitivity to wind speed was observed for videos of structures not included in the training data set (right panel). Figure adapted from Ref. 77.

Because this purely data-driven, machine learning approach has limited capacity for extrapolation beyond the training data distribution[77], it was unable to perform similarly accurate predictions using videos of tree specimens or flag types different from those in the training data. Hence, a generalizable version of visual anemometry in this case, i.e., a method that can make accurate measurements for a diversity of vegetation and engineered structures, would likely require training on a much more comprehensive set of videos and companion anemometer data. Brute-force efforts of this type have proven successful in the past, e.g., ImageNet[80] and COCO[81]. However, developing the equivalent data set for visual anemometry would likely require leveraging a combination of existing open-source data and new, dedicated measurement campaigns. We revisit this possibility in the conclusion of this review.

III. Outlook: Unsolved Challenges, Untapped Opportunities

III. A. Theoretical constraints on visual anemometry

Generalized visual anemometry—a technique that does not require calibration measurements or *a priori* collection of training data—will depend on the discovery of new physical principles that manifest in predictable ways across a diversity of environmental structures exposed to wind. In pursuit of fundamental concepts of this type, we have recently conducted an extensive campaign of concurrent wind and visual measurements in a large-scale wind tunnel[82]. This facility enables controlled studies of selected vegetation with a diversity of morphologies, ranging from grasses to trees (Figure 3).

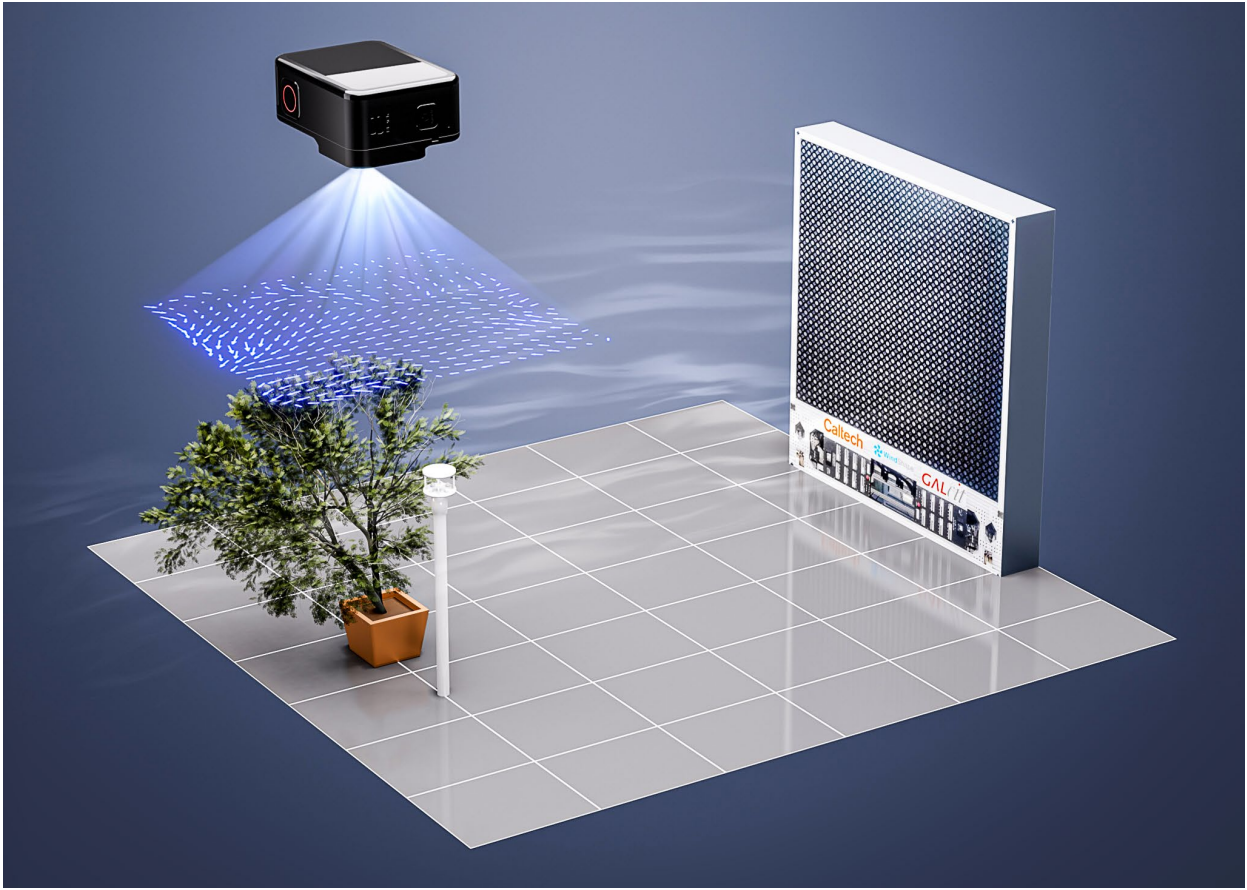

**Figure 3.** Schematic of large-scale wind tunnel measurements of vegetation under controlled wind conditions. A 3-m x 3-m array of 1296 individually addressable fans generates wind conditions with user-defined spatiotemporal profiles and mean speeds up to 20 m/s (right). Vegetation exposed to the wind is recorded from above using a high-speed camera. Spatial cross-correlation of successive images reveals the local, instantaneous displacement of the vegetation, illustrated in the planar vector field. Adjacent to the vegetation, a sonic anemometer (white) measures the local wind speed for comparison to visual anemometry.

The incident wind speed can be described by a two-parameter Weibull probability density function [83,84]:

$$p(U) = \begin{cases} \frac{C_2}{C_1}\left(\frac{U}{C_1}\right)^{C_2-1} e^{-\left(\frac{U}{C_1}\right)^{C_2}}, & U \geq 0 \\ 0, & U < 0 \end{cases} \quad (19)$$

where $C_1$ is a positive-valued, dimensional scale factor that increases for distributions $p(U)$ with higher variance. The dimensionless shape factor $C_2$ typically takes values between 1 and 3 for wind distributions, with values closer to 1 indicating right-skewness of the distribution[83]. Moments of the Weibull distribution can be expressed in terms of $C_1$ and $C_2$; for example, the mean wind speed is given by,

$$\bar{U} = C_1 \Gamma\left(1 + \frac{1}{C_2}\right) \quad (20)$$

where $\Gamma$ is the Gamma function.

The motion of the vegetation can be similarly quantified using the Weibull distribution. Cross-correlation of successive images of the moving canopy creates a displacement vector map[85] representing the spatial distribution of motion induced by the incident wind (Figure 3). Quantile-quantile[84] analysis confirmed that the time-series of the spatially averaged canopy motion can be reasonably approximated by a Weibull distribution with its own scale and shape factors, $C_1^{canopy}$ and $C_2^{canopy}$, respectively.

The dependence of the canopy scale and shape factor on the corresponding wind factors may provide a framework to achieve generalizable visual anemometry. For example, Figure 4 shows that the various vegetation studied to date all exhibit a similar, sigmoidal dependence of the canopy scale factor, $\tilde{C}_1^{canopy}$ on the wind scale factor, $\tilde{C}_1$, where the tilde denotes a vegetation-specific normalization based on the width, height, and center of each sigmoid curve[82]. The physical interpretation of this apparently 'universal' curve shape can be understood by recalling the scale factor as a surrogate for the mean speed of the wind and canopy. At relatively low wind speeds, the dynamic pressure exerted by the wind on the canopy elements may be insufficient to overcome the inertia and elastic restoring force of the structures exposed to wind. In this regime, the slope of the curve in Figure 4 is expected to be nearly zero. At sufficiently high wind speeds, the resistance of the canopy to motion is overcome, and further increases in wind speed correspond to a proportional increase in canopy motion (i.e., the region of linear slope in Figure 4). At high wind speeds, further deflection of the canopy structures is limited by the fixed position of the vegetation roots in the substrate below. This constraint is reflected in the plateau of $\tilde{C}_1^{canopy}$ at large values of normalized wind scale factor $\tilde{C}_1$ in Figure 4.

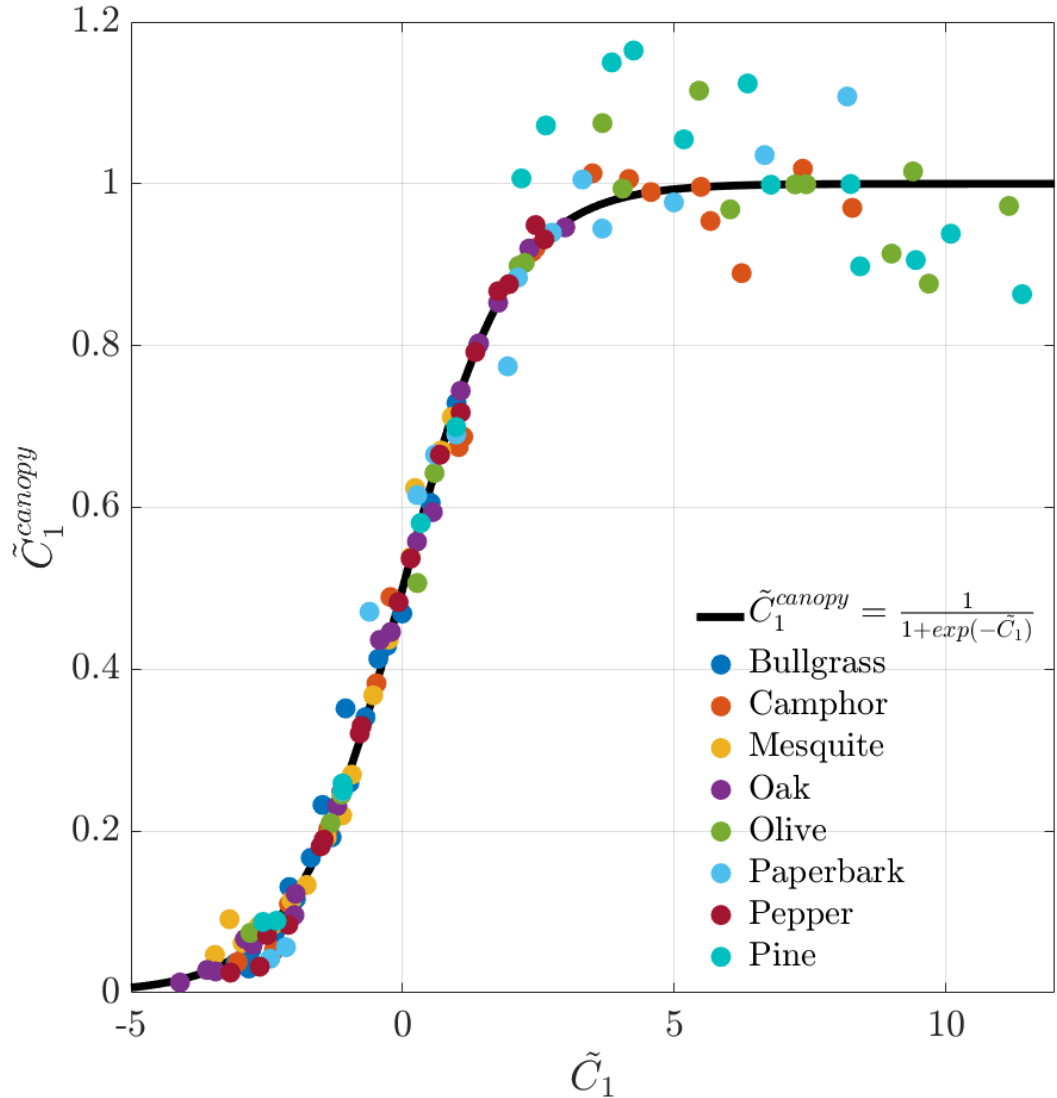

**Figure 4.** Compilation of visual anemometry measurements of eight vegetation species, plotted in terms of the normalized canopy scale factor, $\tilde{C}_1^{canopy}$ versus normalized wind scale factor, $\tilde{C}_1$. Black curve corresponds to sigmoid equation. Data derived from Ref. 82.

A key implication of the sigmoidal response curve is that its slope—a measure of the sensitivity of canopy motion to changes in the incident wind speed—exhibits regimes at both low and high winds wherein visual anemometry may be fundamentally challenged by the lack of a distinct structural response to wind dynamics. Where the response curve has zero slope, it is not possible to accomplish visual anemometry based on the curve of canopy scale versus wind scale. The location and width of each of the aforementioned regions as a function of the dimensional wind speed (e.g., in m/s) is a characteristic of each vegetation type. Placement of a given vegetation type onto the universal curve in Figure 4 required *a priori* knowledge of the incident wind corresponding to each canopy measurement. Generalized visual anemometry would therefore require a means to predict the placement of a given structure onto the universal curve. Additional information based on the shape factor of the motion distribution $\left(\tilde{C}_2^{canopy}\right)$, the visual appearance of the structure, comparison with nearby, similar structures in a canopy (cf. section II.B.1.ii above), fine-scale changes such as leaf reconfiguration (cf. Section II.A.2), or other statistical priors could be useful for achieving this goal. Two physically-motivated priors related to the wind itself include the expected wind speed distribution as a Weibull probability distribution function, and diurnal and seasonal variations in wind that can be known a priori for a given geographic location. Deep learning models that incorporate such physics-based constraints have been widely used in recent years[86].

In the regime of high winds, prior work has also observed that canopy motions can be limited by reaching the maximum physical displacement of the structure[47]. Importantly, the absolute turbulence level (e.g., in m/s) is also highest in the high-wind regime. Hence, measurements of the time-averaged canopy motion (e.g., Figure 4) could be complemented by consideration of the temporal fluctuations in canopy motion in order to disambiguate the kinetic energy trends in high wind-speed conditions.

The spatial scale of measurements in the wind tunnel campaign is limited by the size of the individual trees that could be tested in that facility. A wider range of spatiotemporal scales present in the flow through larger canopies can facilitate the incorporation of additional dynamics, e.g., the "honami" discussed in section II.B.1.ii, to further constrain estimates of the wind incident on a canopy.

We conclude with a prospective discussion of three research avenues that could accomplish the necessary model closure for generalized visual anemometry.

III. B.  New data sources

A data-driven approach to generalized visual anemometry could use the discovered universal curve as a statistical prior in a physics-informed machine learning framework. This approach anticipates that measurements collected without ground-truth wind speed measurements should exhibit a scale factor relationship between the wind and canopy distributions that is sigmoidal, as in Figure 4. The relationships between the wind and canopy shape factors may provide additional physical constraints to enable a data-driven model that can extrapolate beyond its training data set.

To be sure, this approach does not obviate the need for comprehensive data collection to train the neural networks or other machine-learning representations of the underlying physics. However, there exists a growing set of data sources that could be leveraged for this purpose. These include open-source, near-ground imagery[87], e.g., from long-term ecological measurement campaigns[88–93], hazard monitoring systems such as those deployed for wildfire detection[94–102], and in the built environment, traffic and security cameras[103,104]. A large number of existing meteorological measurement campaigns could also be augmented with concurrent video collection to provide large volumes of new labeled data to train machine learning models[105–115].

Emerging commercial satellite data feeds can provide a potentially transformative data source if extended to time-resolved imagery, as the wide area coverage and frequent revisits of remote locations provide data that is inaccessible by other means[99,116]. Although the distant vantage of satellite data can limit the spatial and temporal resolution of near-ground canopy measurements, recent advances in AI-based image upscaling could enable features of the canopy essential for visual anemometry to be recovered from low-resolution data following initial training from benchmark data sets[117].

III. C.  New computational tools

The two primary approaches toward visual anemometry that have been explored in this review—first-principles physical modeling and data-driven machine learning—have both been considered thus far from a perspective depending on empirical measurements of the relevant flow-structure interactions. Advances in high-performance computing now make it feasible to achieve physically realistic computational simulations of wind interactions with geometrically complex structures such as vegetation[118–120]. Hence, another promising route to generalized visual anemometry could leverage simulations to complement the aforementioned field measurement campaigns. Numerical simulations provide the added benefit of enabling complex details of environmental structures, e.g., the branches of a tree, to be tracked with high spatiotemporal fidelity. Because imagery from cameras provides only a two-dimensional projection of the three-dimensional structure kinematics, the set of parameters used to describe the canopy is limited to quantities derived from that projection. The canopy motion determined from image cross-correlation is one example (Figure 3). By contrast, numerical simulations could provide three-dimensional kinematic data, from which a richer set of physical descriptors could be derived to quantify the canopy response to incident wind. That higher-dimensional description can better delineate different modes of structural response and could also be used for the task of identifying and classifying structures of interest in a machine learning context. Accurate simulations of wind response can also be leveraged in virtual reality and gaming contexts, which can potentially engage a broader audience in efforts to crowdsource measurements for visual anemometry training data[121].

III. D.  New physics

The ultimate solution to the challenge of generalized visual anemometry may lie in a combined strategy that leverages knowledge of canonical flow-structure interactions, such as those introduced in section II. A, along with libraries of representative wind interactions from empirical observations and from analogous computational models. However, the most exciting role for the physics community may lie in a third approach: the discovery and development of new physics concepts that augment our current knowledge of the nature of flow-structure interactions as well as our remote sensing capabilities.

Although the flow-structure interactions to be exploited by visual anemometry are a manifestation of classical mechanics—a subfield of physics that is ostensibly mature in comparison to, say, quantum science—knowledge of those physics is still limited to a relatively small set of simplified geometries. The appeal in section II. A to objects with circular cross-sections, slender or planar geometries, and moderate elasticity was by necessity, as established models for the physics of flow structure-interactions have not evolved beyond those relatively simple configurations despite intensive study for more than a century[122–131]. A historical limitation on the study of flow-induced motion of more complex structures was the inability to visualize the associated fluid-solid interactions with high spatiotemporal resolution. However, the advent of high-speed laser velocimetry[132,133], 3D flow tomography[128,134], and algorithms to compute the pressure field corresponding to flow velocity measurements[135,136] now make it possible for experimental physicists to revisit the classical mechanics in geometric configurations approaching the complexity of structures relevant for visual anemometry. Indeed, important new results have emerged in the past few years, from plant-scale to canopy-scale, which have improved our understanding of flow-structure interactions and which bring the present goal of generalized visual anemometry closer to realization[70,137].

Modern model reduction techniques from dynamical systems theory[138–140] have the potential to distill high-dimensional datasets such as those derived from the aforementioned new experimental measurements. Physicists familiar with the challenge of dimensionality reduction in other areas of nonlinear dynamics could apply many of the same tools here. Simplified kinematic motifs of the observed structure motion, extracted using model reduction, may prove to be robust correlates of the incident wind speed and direction. That could also provide a target for unsupervised machine learning approaches that aim to classify or even deduce material properties of objects in the wind based solely on their observed motion.

While the concept of visual anemometry took initial inspiration from our human powers of visual observation, the spectrum of visible light represents a relatively small band of the electromagnetic radiation that is absorbed, reflected, and emitted by both natural and engineered objects that could be used for visual anemometry. The range of applications of the concepts introduced here could be further expanded by physicists interested in exploring the broader spectrum of electromagnetic radiation associated with objects covering the Earth's surface that are subjected to local winds. An immediate example is infrared radiation, which could enable visual anemometry measurements at night. Imaging at longer wavelengths, e.g., millimeter-wave imaging[141–143], could also potentially be used to circumvent optical interference such as cloud cover, provided that the spatial resolution of those measurements still enables structural motions to be resolved. Additional optical properties, such as the polarization of reflected light, could be used to infer not only translational motion of objects but also changes in object orientation associated with flow-induced bending and torsion of reflective surfaces[144] such as leaves and blades of grass. These signatures could provide additional means to discriminate between regimes of flow speed and direction incident on the objects.

Finally, it is important to recall that 70 percent of the Earth's surface is covered by water. Inference of wind fields near the ocean surface is confounded by the more complex deformations associated with the air-water interface[144–147]. This presents challenges but also opportunities. For example, the high-wind plateau in structural response observed for ground-mounted structures (e.g. Figure 4) need not limit correlations between ocean surface deformation and wind speed in similarly high-wind regimes. Hence, a larger range of wind speeds may be accessible to visual anemometry over the ocean as compared to the method applied on land. In addition, ocean measurements can potentially leverage not only the kinematics of the air-water interface, but also the wind-induced motion of ocean spray above the surface[148] and the water-induced motion of submerged vegetation[126,127,149].

The aforementioned list is merely illustrative of avenues for new contributions from the physics community. If this discussion has been successful, the opportunities and challenges associated with visual anemometry that have been introduced herein will encourage the reader to pursue one or more of these research directions. Visual anemometry provides a unique opportunity for the physics community to contribute to a variety of important and far-reaching topics in global sustainability.

Text Box 1

> **Applications of Visual Anemometry**
>
> *Renewable energy*
>
> To achieve net-zero greenhouse gas emissions in the U.S. by 2050, it is estimated that increases in wind and solar capacity of 6-28x and 9-39x, respectively, are required[150]. Similarly large increases in renewable energy are needed to meet global decarbonization targets. This unprecedented scale-up will require widespread proliferation of wind and solar generation in geographic regions where renewables have not previously been sited[20,151]. Diversified siting of renewable energy infrastructure creates two main challenges related to wind measurements. First, wind patterns are more uncertain in new locations that have shorter historical observation records. Second, new sites may have lower quality wind, with characteristics that are more difficult to incorporate into existing forecasting methods, such as terrain complexity[105,106,152,153] and variable land use (e.g., urban environments).
>
> Historically, two parallel approaches have been used for wind field estimation for renewable energy resource assessment. First, numerical weather forecasting models predict the winds in the atmosphere based on an approximate form of the governing equations[154]. These models require parameterizations to represent complex processes that cannot be directly resolved with available computing resources, such as turbulence, convection, and clouds. While numerical weather models provide detailed spatial and temporal coverage, the approximations in the models result in significant uncertainties in wind forecasts[155], especially near Earth's surface where turbulence is higher than aloft. In the second approach, *in situ* sensors are used to observe the wind with few, if any, assumptions required regarding the nature of the wind dynamics[33,105,156,157]. Yet these sensors are both relatively high cost and lack spatial coverage. At the intersection of these parallel approaches, data assimilation is used to combine *in situ* observations with numerical models[158–160], but uncertainties remain in locations not covered by the measurements. Visual anemometry can provide a third, complementary approach to wind field estimation with characteristics similar to *in situ* observations, but with higher spatial coverage. A promising avenue may also leverage wind estimates from visual anemometry for data assimilation.

Since most physical objects to be used for visual anemometry, both natural and engineered, exist tens of meters or closer to Earth's surface, measurements via visual anemometry directly quantify winds near this nominal height (Figure 5). Small-scale wind generators, such as recently developed vertical-axis wind turbines, are designed with hub-heights on the order of ten meters [161], making visual anemometry measurements directly applicable to the design of those systems. However, utility-scale horizontal axis wind turbines operate at hub-heights between 50-200 meters, and with rotor diameters 80-300 meters in size. For wind measurements made through visual anemometry to be used to estimate the winds incident to these utility-scale horizontal axis turbines, model-based extrapolation methods must be used[162]. Wind extrapolation to heights above a given measurement location is common, as typical weather stations also provide wind measurements 10 meters above Earth's surface and surface winds reported by typical weather and climate models are also at 10 meters.

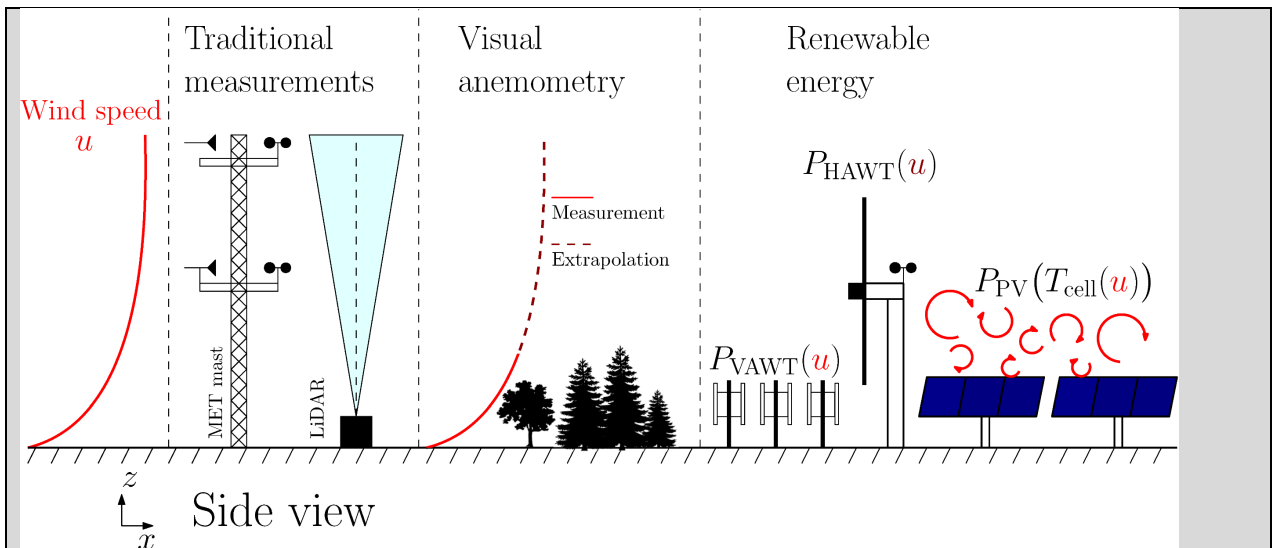

**Figure 5.** Traditional tools for wind measurement (left) include meteorological towers carrying sensors for wind speed and direction, as well as light detection and ranging (LiDAR) systems. Visual anemometry (center) measures near-ground wind conditions relevant to solar photovoltaic (PV) and solar thermal energy infrastructure, as well as smaller wind turbines (e.g., vertical-axis wind turbines, VAWT). The measurements can also be extrapolated to higher altitudes relevant to traditional horizontal-axis wind turbines (HAWT). [Tree graphic adapted from: http://clipart-library.com/free/forest-silhouette-clip-art.html]

Solar power production primarily depends on spatiotemporal variations in irradiance. Irradiance is driven by known deterministic variations, such as seasonal and diurnal cycles, as well as stochastic variations of atmospheric clouds and aerosols that are challenging to predict. Therefore, physics-based irradiance forecasts rely on numerical weather prediction[163]. As noted above in the context of wind energy, visual anemometry may provide a mechanism for improved weather forecasting by increasing the availability of wind flow measurements for data assimilation.

Beyond irradiance, the efficiency of solar photovoltaic (PV) cells is rated at standard test conditions (STC) of one Sun of irradiance at an air-mass ratio of 1.5 (i.e., sunlight passing obliquely through the equivalent of 1.5 times the atmospheric length at zenith) and a cell temperature of 25 degrees C. Yet PV cells typically operate at higher temperatures[164]. Solar PV efficiency decreases by approximately 0.1 to 0.5% per Kelvin above STC, but the magnitude of degradation is cell-specific[165]. To estimate efficiency in field conditions, solar cell manufacturers provide a method to approximate the cell temperature based on an empirical indicator called the nominal operating cell temperature (NOCT). Wind speed is a required input to this cell temperature approximation[166]. Improving wind estimates increases the accuracy of cell temperature and cell efficiency predictions[167]. Finally, emerging research seeks to optimally site and design solar farms to maximize passive convective cooling to reduce cell temperature[168]. Such methods require site-specific wind estimates[22] which may be provided by visual anemometry.

*Urban flow physics*

People increasingly live in urban environments. In recognition of this important trend, the United Nations Sustainable Development Goal (SDG) 11 focuses on sustainable cities and communities[169]. Air flow in urban environments affects the energy efficiency and resilience of engineered structures, pollution dispersion and air quality, and the future of urban air mobility. Given the broad impacts of urban air flow, and the limited fidelity of present observations and predictive models, urban air flow represents a grand challenge in environmental fluid mechanics[170].

Air flow affects the structural resilience and energy efficiency of buildings. Design standards incorporate site-specific wind characteristics, including extreme wind gusts[171], which are difficult to measure or numerically model. Urban airflow also affects thermal convection in cities, which impacts the energy consumption of building heating and cooling systems, and the effectiveness of natural ventilation[172]).

Finally, the design of future aircraft and airspace for urban air mobility depends on our ability to predict the turbulent flow within and around urban environments[173]. Safe and reliable transport of people and goods requires detailed knowledge of wind gusts and turbulence, as contemporary control methods have reduced success in navigation and object avoidance in uncertain wind environments[174].

Flows in urban environments are heterogeneous and complex; these traits reduce the accuracy of numerical flow predictions and severely limit the accuracy of spatially extrapolated pointwise flow field measurements. There is an urgent need for increased spatiotemporal coverage of urban air flow measurements[170] for both validation and uncertainty quantification of numerical models[175,176] as well as for data assimilation[177]. Visual anemometry can provide a new paradigm for urban wind field sensing with wide spatial coverage and high spatiotemporal resolution (Figure 6).

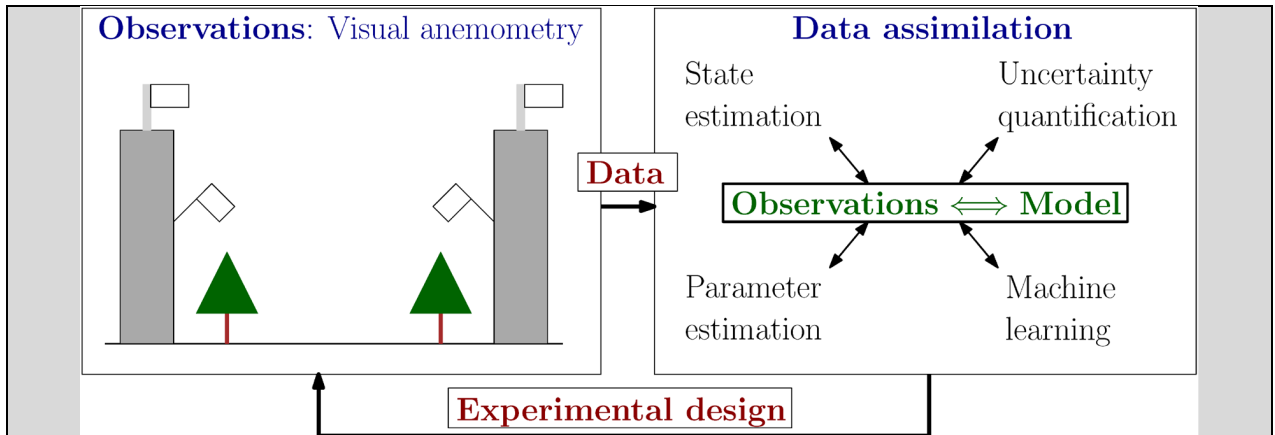

**Figure 6.** Schematic of visual anemometry for urban flow physics. Observations of the motion of naturally occurring objects such as trees and engineering objects such as flags can potentially be used in a data assimilation framework to inform wind state estimation, model parameter estimation, and uncertainty quantification, and machine learning of wind models. Conversely, developed models can inform optimal placement of additional passive sensors.

*Environmental and ecological processes*

Transport, mixing, and atmospheric conditions driven by the wind are central to numerous environmental and ecological processes (Figure 7). Wildfire prediction and mitigation are notable applications where detailed wind mapping can play a critical role. In the United States alone, wildfires have cost an average of $13.4B USD/yr[178,179] in damages over the last five years and are predicted to become even more prevalent as warmer and drier conditions driven by climate change lead to more protracted and active fire seasons[180,181]. While temperature, humidity, and stability conditions are critical contributors to the intensity of a wildfire, wind conditions play a leading role in determining the speed and direction of the wildfire spread[182–184]. In chaparral ecosystems such as coastal Southern California, the regions most exposed to extreme wind events (e.g., the katabatic Santa Ana winds) have been linked to increased fire danger and larger fires[185,186].

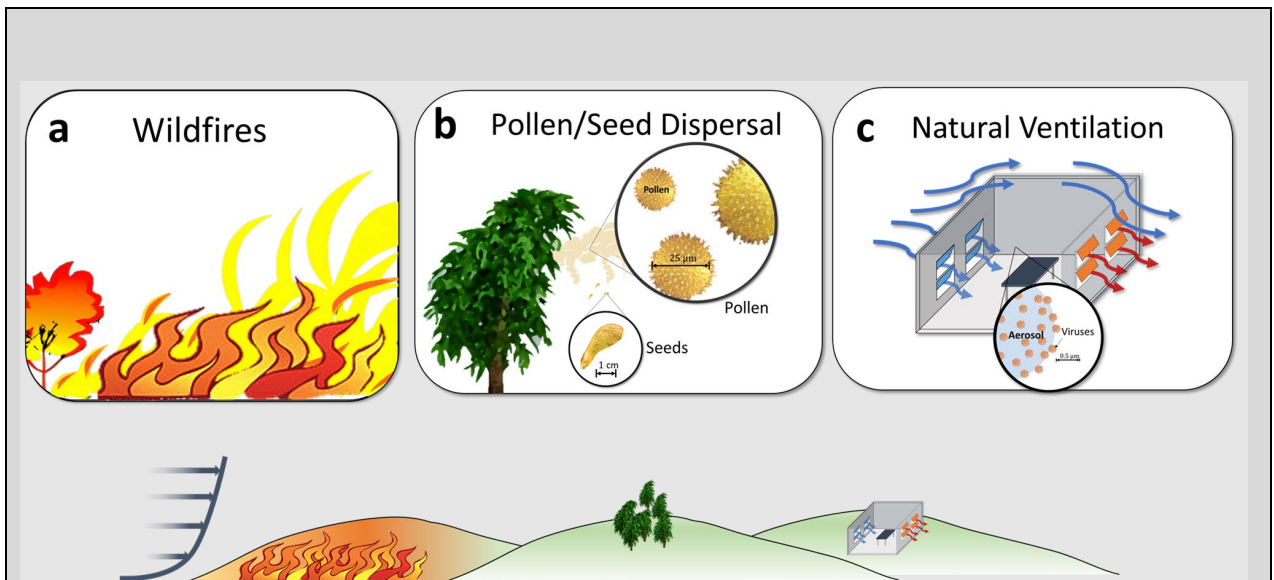

**Figure 7.** A diverse range of ecological processes are mediated by near-ground winds, including (a) wildfires, (b) pollen and seed dispersal, and (c) natural ventilation.

Prevailing wind conditions and their associated turbulence also drive the dispersal of critical scalar quantities (e.g., heat and mass) and particulates central to various ecological and environmental processes. Long distance dispersal of seeds[14], spores[187], and pollen[188], whether by wind or organisms, is a critical yet poorly understood survival strategy[11,14,189,190] for species that adapt to changing habitats by overcoming geographic isolation[10,191,192]. Many of the world's most important agricultural grains, including wheat, barley, corn, and rice, are grasses that are pollinated primarily through the wind due to their lack of flowering structures[193]. Though such a mechanism is most prevalent for plants located in close proximity, e.g., within a range of 1 km[194], airborne transport and mating of taller plants, e.g., trees, has been documented at distances exceeding 10 km [12,188,195]. Understanding these dynamics is critical for ensuring or minimizing cross-pollination between various crops; the latter is especially critical to limit contamination from genetically modified crops[194] in the natural environment. Even propagated material that is transported by organisms depends in part on wind dynamics for its dispersal. The efficacy of crop treatment by pesticides is also directly impacted by local wind conditions[196–201]. Current guidelines reflect an inability to precisely quantify the wind flows that advect pesticide chemicals, presenting another application of visual anemometry with broad impact.

Natural dispersal, convection, and mixing by wind has similarly been leveraged by humans for various engineering purposes, including for pollution dispersal[202] and natural ventilation[203,204]. Proper ventilation is necessary to ensure a healthy indoor environment, but it comes with a tangible energy cost. Ventilation comprises approximately 11% of the nearly 7 quadrillion BTU (2 million GWh) and $141 B USB spent by commercial buildings in the US alone[205]. This expenditure does not include the energy spent on space heating or cooling, which are both also substantial (i.e., 32% and 8% of energy consumption by commercial buildings in the US, respectively). Natural ventilation presents an efficient and largely passive alternative to conventional ventilation approaches. This approach uses naturally-occurring forcing from wind and/or buoyancy to exchange air between the indoors and outdoors through openings in a building structure. While this resource is freely available in appropriate climates, it can be challenging to control and predict[206,207] due to the inherent complexity and variability of the airflow inside connected rooms and around the structure[208–210]. Designing a system to adequately take advantage of wind-driven forcing requires detailed knowledge of the turbulent wind patterns in and around the building across diurnal and seasonal variations[211]. This represents another potentially transformative application of visual anemometry.

**Acknowledgements**

The authors gratefully acknowledge seminal contributions from Jennifer L. Cardona in development of several of the concepts presented in this review, as well as discussions with Katie Bouman, Jennifer Sun, Yisong Yue, and Pietro Perona at Caltech. Additional helpful discussions occurred in the CV4Ecology Summer Workshop, supported by the Caltech Resnick Sustainability Institute. Constructive feedback from the anonymous reviewers led to meaningful improvements to the presentation of the material in this manuscript. Funding was generously provided by the National Science Foundation (Grant CBET-2019712) and the Center for Autonomous Systems and Technologies at Caltech. Additional support from Heliogen is gratefully acknowledged.

**The authors declare no competing interests.**